# Probing light primordial black holes through noncold dark matter

Yu-Ming Chen [*]
*Department of Physics, Carleton University, Ottawa, Ontario K1S 5B6, Canada*

We study the matter power spectrum constraint on primordial black holes (PBHs) by the dark matter (DM) emitted through Hawking radiation. We particularly focus on the scenario where a PBH, with mass ranges between 1 g and $10^9$ g, evaporates before big bang nucleosynthesis. In addition to that, we consider the case where PBH abundance is scarce and there is no early PBH domination taking place. On the DM side, we assume a fraction of the population is produced from PBH evaporation, while the remaining part is the regular cold dark matter, which is produced by some genesis processes that decouples later on. Therefore, in the rest of the cosmological history, DM interacts solely through gravity. Under this condition, there is no thermal equilibrium ever established between DM and standard model plasma. An important feature in our analysis is that, for the light PBH we consider, its temperature is much larger than the mass of DM, which is consequently produced ultrarelativistically and requires a protracted time to become matterlike. In this context, even though a PBH evaporates in the very early Universe, PBH-produced DM could still be energetic and smooth out the small scale structure at a much later time. By the precision measurement on the matter power spectrum from cosmic surveys, we are able to set a joint constraint on light PBHs and the noncold DMs they produced.

## I. INTRODUCTION

The standard model (SM) in particle physics has been hugely successful for over 50 years. However, it remains fundamentally incomplete. A mystery has lingered around even longer: the identity of the dark matter (DM) [1]. We now have copious evidence, from rotational curves [2,3], gravitational lensing [4–7] and bullet clusters [8–10], about the existence of DM. Yet we remain ignorant about its properties. It is important to note that all of these evidences are purely gravitational as all the direct detection experiments on the ground have yielded null results [11–13]. Therefore the only thing we do know is that, DM interacts with SM particles through gravitational interaction and in the most pessimistic scenario this might be the exclusive way to probe DM.

Since we barely know any properties of DM, including its mass, spin, and interaction type, the potential candidates of DM span a wide landscape of mass parameter space. One of the popular candidates of DM is the primordial black hole (PBH) [14–16]. PBH formation in the early Universe could result from inhomogeneities [17–19], early matter domination [20–22], or phase transition [23–25]. However, due to Hawking radiation [26], PBHs with mass $\lesssim 10^{15}$ g would have been evaporated by now if there is no memory burden effect [27]. As a consequence, apart from big-bang nucleosynthesis (BBN), which constrains $10^9\ \mathrm{g} \lesssim M_{\mathrm{PBH},0} \lesssim 10^{13}$ g [28–31], and isocurvature [32], the parameter space for light PBHs with mass $\lesssim 10^9$ g remains largely unconstrained. In this work, we combine two concepts above where a fraction of DM population is from some nonthermal production mechanisms, such as inflaton decay [33], topological defects [34,35], or even purely gravitational effect [36], on the other hand, the remaining part of the DM is coming from the Hawking radiation of light PBHs, which constitute the so-called fractional noncold DM (f-NCDM) scenario. As a proof of concept, the temperature of PBH could be determined semiclassically[1]

$$T_{\mathrm{PBH}} = \frac{M_{\mathrm{pl}}^2}{8\pi M_{\mathrm{PBH}}}, \tag{1.1}$$

where $M_{\mathrm{pl}} = 1.2 \times 10^{19}$ GeV is the Planck mass. Take a $10^9$ g PBH for example, its temperature is about $10^4$ GeV. If we approximate the energy of DM to the temperature of the PBH, we see that DMs from the PBH are produced ultrarelativistically ($E_{\mathrm{DM}} \gg m_{\mathrm{DM}}$) even for weak scale DMs. On the other hand, $10^9$ g PBH evaporates when the temperature of the Universe is about 15 MeV, as a consequence, DMs from PBHs are not only relativistic but also hot ($E_{\mathrm{DM}} \gg T_\gamma$). Here we consider the case where DMs only interact gravitationally. By this assumption, DMs have never established thermal contact with the SM plasma and can only lose their energy as the Universe expands. Even just a small fraction of DMs arises from PBH evaporation, if it is still energetic when recombination happens, it could leave observable imprints on the cosmic microwave background (CMB) and the structure formation, offering unique probes of both particle physics and early Universe cosmology.

[*]Contact author: yumingchen@cmail.carleton.ca

[1]Throughout this paper, we use the natural unit, $\hbar = c = k_B = 1$.

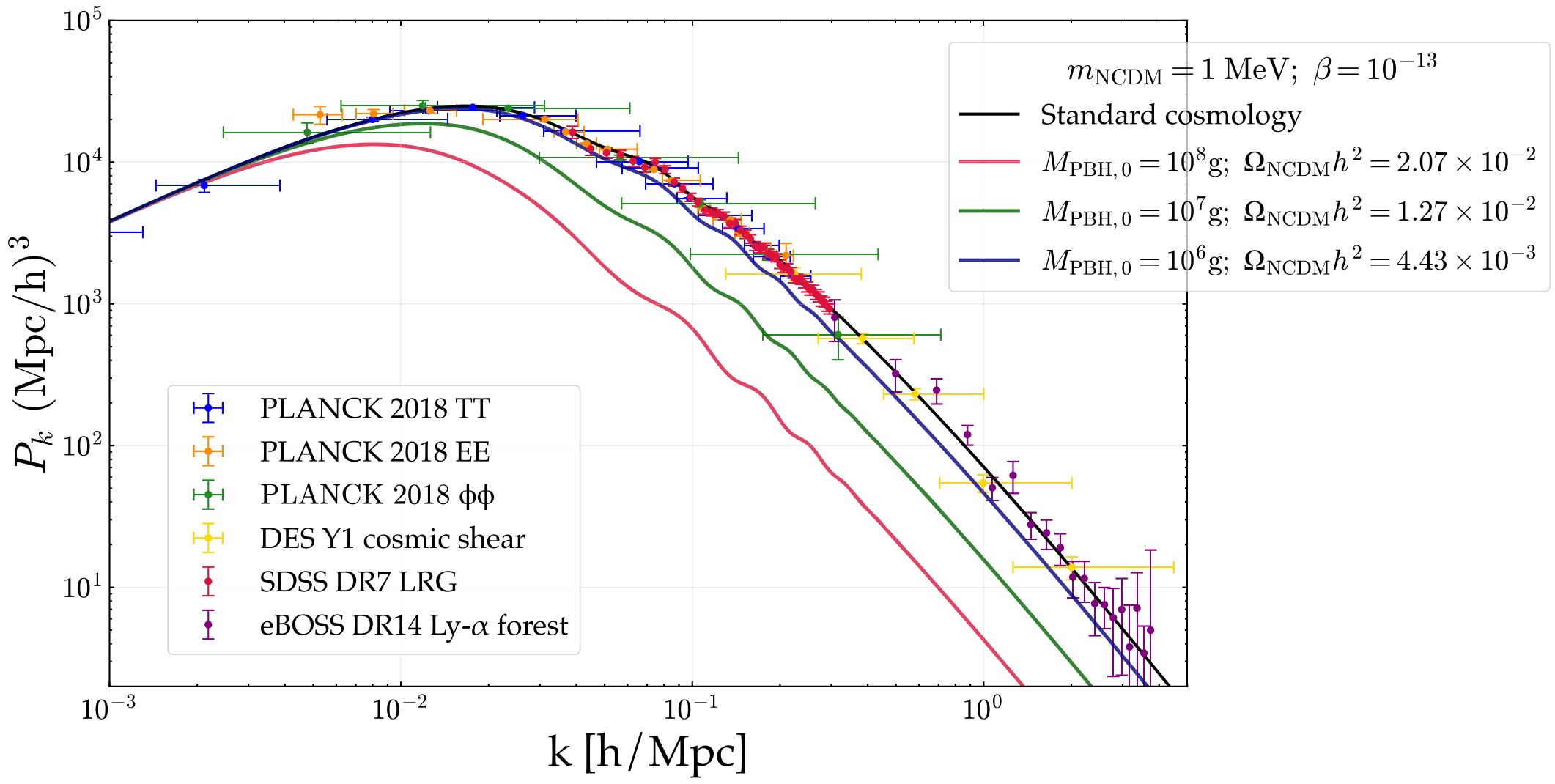

FIG. 1. Matter power spectrum with the NCDM mass $m_{\mathrm{NCDM}} = 1$ MeV and the parameter $\beta = 10^{-13}$, which determines the abundance of PBHs in Eq. (2.2). NCDM produced by PBH before BBN could still be warm at the time of structure formation and its free-stream effect smooth out the structures and suppress the matter power spectrum.

In the standard scene of structure formation, small density fluctuations start to grow after horizon crossing and CDMs, lacking photon pressure, serve as seeds for small scale structures to form first then clump up to form larger structures in the so-called "bottom-up" approach. However, NCDMs suppress small-scale perturbations through free streaming [37,38] and several studies have put constraints on its fraction in the total DM relic density [39–44]. Here we also consider the suppression of the matter power spectrum on a large wavenumber, but we take a step back to put a constraint on the abundance of PBHs as being the source of NCDM. The free stream effect is clearly encoded in the large-$k$ part of the matter power spectrum as shown in Fig. 1. The data points with error bars are taken from the Planck satellite measurement on CMB temperature and polarization angular power spectra [45], dark energy survey (DES) year 1 results on cosmic shear [46], sloan digital sky survey (SDSS) data release 7 (DR7) on luminous red galaxies [47], and the Lyman-$\alpha$ forest, which observes high redshift quasars to measure the baryon acoustic oscillation peak [48]. The black line, which highly matches the data, represents the success of the standard ΛCDM model. We vary the PBH mass while fixing its initial energy density and the NCDM mass. The suppression of the power spectrum is obvious and thanks to the precise measurements, we are able to set the most stringent constraint on PBHs with masses in the range $1\ \mathrm{g} \leq M_{\mathrm{PBH,0}} \leq 10^9$ g, by the suppression caused by the PBH-produced NCDM.

The paper is organized as follows: Sec. II describes the basic properties of a Schwarzschild PBH and places a constraint on its initial energy density by requiring that there is no early matter domination. Section III then describes the DM produced by PBH, including how to correctly obtain its phase space distribution (PSD), number density as well as energy density. From here we can give another constraint on PBH abundance by the overproduction of DM. In Sec. IV, we place constraint on PBH parameter space and discuss the physics therein, especially the interplay between the PBH mass and DM mass. We also comment on the case where the PBH dominates the universe prior to BBN. Finally, we summarize the paper in Sec. V.

## II. PRIMORDIAL BLACK HOLE IN THE EARLY UNIVERSE

### A. PBH production

We first address the production of PBHs in the early Universe. PBHs considered here are Schwarzschild black holes with no spin and zero charge. We consider critical collapse as the formation mechanism where the PBHs are formed when the overdensity regions recollapse after entering the horizon. The fluctuation in a Hubble patch should be larger than Jean's scale, which is $\sqrt{\omega}$ times the horizon size for equation of state $p = \omega\rho$ [17]. In the radiation dominated Universe, we can relate the initial mass of PBH to the time $t_i$ (or equivalently the temperature $T_i$) when it was formed by the above argument

$$M_{\mathrm{PBH},0} = \frac{4\pi}{3}\left(\frac{\sqrt{\omega}}{\mathcal{H}(T_i)}\right)^3 \rho_r(T_i), \tag{2.1}$$

where $\omega = \frac{1}{3}$ for a radiation dominated Universe, $\mathcal{H}$ is the Hubble parameter, and $\rho_r$ is the radiation energy density. We note in passing that PBHs formed in the early Universe could have a wider mass distribution [49,50]. However, for the mass and the critical collapse mechanism we consider here, the effect has no significant deviation from the one considering monochromatic distribution. Therefore, in the following text, we consider PBHs to be monochromatic for simplicity.

The initial abundance of PBHs is parametrized with a parameter $\beta$, which is defined as the energy density ratio between PBH and all radiationlike species when PBHs were formed

$$\beta = M_{\mathrm{PBH},0}\frac{n_{\mathrm{PBH}}(T_i)}{\rho_r(T_i)}, \tag{2.2}$$

where it can be related to $\Omega_{\mathrm{PBH}}$ today as

$$\Omega_{\mathrm{PBH}} = \beta\Omega_\gamma\left(\frac{g_{*S}(T_i)}{g_{*S}(T_{\mathrm{CMB}})}\right)^{\frac{1}{3}}\frac{T_i}{T_{\mathrm{CMB}}} \tag{2.3}$$

in the assumption of no evaporation or accretion.

### B. PBH evaporation

In real life, PBH evaporates through Hawking radiation. The mass loss rate of PBH is given by the Stefan-Boltzmann law

$$\frac{dM_{\mathrm{PBH}}}{dt} = -4\pi r_S^2\cdot\mathcal{G}\sigma T_{\mathrm{PBH}}^4 = -\frac{\mathcal{G}g_*^{\mathrm{PBH}}M_{\mathrm{pl}}^4}{30720\pi M_{\mathrm{PBH}}^2}, \tag{2.4}$$

where $\mathcal{G}\approx 3.8$ is the appropriate graybody factor [followed by Eq. (5) in [51] ], $\sigma = g_*^{\mathrm{PBH}}\pi^2/120$ is the Stefan-Boltzmann constant in which $g_*^{\mathrm{PBH}}$ takes into account all the degrees of freedom of particles that PBH emits, and $r_S = 2M_{\mathrm{PBH},0}/M_{\mathrm{pl}}^2$ is the Schwarzschild radius. By integrating Eq. (2.4) from 0 to the time when all of its mass completely evaporates away, we get the lifetime of PBH

$$\tau_{\mathrm{PBH}} = \frac{10240\pi}{\mathcal{G}g_*^{\mathrm{PBH}}M_{\mathrm{pl}}^4}M_{\mathrm{PBH},0}^3. \tag{2.5}$$

Note that in principle the lower limit of integration should start from $t_i$, the time when PBH was formed, however, we argue that $t_i$ is very close to zero. Indeed, for a $10^9$ g PBH, the heaviest case in our consideration, the lifetime is about 0.41 sec. However, the time of formation is only $2.38\times10^{-29}$ sec after the big bang, which is negligible and justifies our approximation. Moreover, Eq. (2.5) can be used to derive the photon temperature when PBH evaporates. For a radiation dominated Universe,

$$\tau_{\mathrm{PBH}} = \frac{1}{2\mathcal{H}(T_{\mathrm{eva}})}$$
$$\Rightarrow T_{\mathrm{eva}}(M_{\mathrm{PBH},0}) \simeq 23.01\ \mathrm{GeV}\left(\frac{88.92}{g_*(T_{\mathrm{eva}})}\right)^{\frac{1}{4}}\left(\frac{10^6\ \mathrm{g}}{M_{\mathrm{PBH},0}}\right)^{\frac{3}{2}}, \tag{2.6}$$

where 88.92 is the value of $g_*$ at the time when a $10^6$ g PBH evaporates. One should be careful that although the $g_*$ effect is negligible for PBH mass $\lesssim 10^7$ g, it will yield a factor of 2 difference for heavier PBH with evaporation temperature smaller than QCD scale. Later we will see the $g_*$ effect manifests itself as a relaxation of constraint in PBH parameter space. In addition to the lifetime, we can derive the PBH mass evolution from Eq. (2.4)

$$M_{\mathrm{PBH}}(t) = M_{\mathrm{PBH},0}\left(1-\frac{t-t_i}{\tau_{\mathrm{PBH}}}\right)^{\frac{1}{3}}. \tag{2.7}$$

Since PBH is always nonrelativistic, its energy density redshifts as $a^{-3}$, slower than that of the radiation, which redshifts as $a^{-4}$. In this work, we focus on the case where there is no early PBH domination, and since PBH is a subdominant component of the universe, we also neglect the reheating effect when evaporating. Correspondingly, we define a critical $\beta$ on PBH abundance at the time when PBH is about to evaporate

$$1 > \frac{M_{\mathrm{PBH},0}n_{\mathrm{PBH}}(T_{\mathrm{eva}})}{\rho_r(T_{\mathrm{eva}})} = \frac{M_{\mathrm{PBH},0}n_{\mathrm{PBH}}(T_i)\left(\frac{T_{\mathrm{eva}}}{T_i}\right)^3}{\rho_r(T_i)\left(\frac{g_{*S}(T_{\mathrm{eva}})}{g_{*S}(T_i)}\right)^{\frac{4}{3}}\left(\frac{T_{\mathrm{eva}}}{T_i}\right)^4}$$
$$= \beta\left(\frac{g_{*S}(T_i)}{g_{*S}(T_{\mathrm{eva}})}\right)^{\frac{4}{3}}\frac{T_i}{T_{\mathrm{eva}}}. \tag{2.8}$$

We make use of Eqs. (2.1) and (2.6) to get the expressions of $T_i$ and $T_{\mathrm{eva}}$, respectively. We place a *soft* upper bound on $\beta$

$$\beta < \beta_c = \left(\frac{g_{*S}(T_{\mathrm{eva}})}{g_{*S}(T_i)}\right)^{\frac{4}{3}}\frac{T_{\mathrm{eva}}}{T_i}$$
$$\simeq 6.41\times10^{-12}\left(\frac{g_*(T_{\mathrm{eva}})}{g_*(T_i)}\right)^{\frac{13}{12}}\left(\frac{10^6\ \mathrm{g}}{M_{\mathrm{PBH},0}}\right), \tag{2.9}$$

where $\beta_c$ is the critical value for not having an early PBH domination and in the early Universe, we approximate $g_*\approx g_{*S}$. We use the fact that the PBH temperature is always much larger than the mass of every SM particle and the fermionic DM we considered, therefore $g_*^{\rm PBH} = 108.75$ is fixed. Of course, the assumption of no PBH domination is not mandatory as we do not have direct evidence about which species run the Universe before the BBN. In fact, reheating by inflaton decay suggests that the Universe might experience an early matter domination period prior to the onset of the BBN [52,53]. Despite that, our goal is trying to prove that PBHs do not have to be so abundant to have an impact on the Universe, only a small population would be enough to alter the matter power spectrum and CMB observations. Nevertheless, we comment on the scenario of early PBH domination in Sec. IV C.

## III. BOOSTED DARK MATTER FROM A PRIMORDIAL BLACK HOLE

Black holes (BHs) emit particles with masses smaller than their temperature in the form of Hawking radiation. The PBHs we consider here have temperature much greater than all the SM particles as well as the NCDM, thus not only all the particles can be produced from PBHs, they are produced ultrarelativistically. As we mentioned earlier, we consider two components of DMs, a fraction of them are coming from the early time gravitational production mechanism, which follows the usual evolution pattern and constitutes a regular CDM; while the remaining part of the DMs are produced by evaporating PBHs and are much more energetic than the former ones. We can make a simple estimation, by relating $E_{\rm NCDM} \sim T_{\rm PBH}$, we get the ratio between the energy of NCDM and the photon at the time of PBH evaporation

$$\frac{E_{\rm DM}}{T_{\rm eva}} \sim 4.59 \times 10^5 \left(\frac{g_*(T_{\rm eva})}{88.92}\right)^{\frac{1}{4}} \left(\frac{M_{\rm PBH,0}}{10^6\ {\rm g}}\right)^{\frac{1}{2}}. \quad (3.1)$$

Since DMs here only have gravitational interaction with SM particles, they cannot efficiently compensate this huge energy difference by depositing energy to the SM sector and therefore need a great period of time to cool off as will be shown in Eq. (3.15).

NCDM would have a PSD different from the usual Fermi-Dirac distribution after diffusing. To acquire the NCDM spectrum, we start from considering its energy density. By definition,

$$d\rho_{\rm NCDM} = g_{\rm NCDM} \int \frac{d^3\vec{p}}{(2\pi)^3} E\, d\mathcal{F}_{\rm NCDM}(x), \quad (3.2)$$

where $x = E_{\rm NCDM}/T$ is conserved for relativistic objects and $\mathcal{F}_{\rm NCDM}$ is the PSD of NCDM.

On the other hand, from the production point of view, PBH only differs from a blackbody by a graybody factor $\mathcal{G} \approx 3.8$ and NCDM, just like ordinary blackbody radiation, should follow the thermal Fermi-Dirac distribution to begin with. Thus the luminosity of NCDM is given by

$$L_{\rm NCDM} = \mathcal{G} \times g_{\rm NCDM} \int \frac{d^3\vec{p}}{(2\pi)^3} \frac{E}{e^{xT/T_{\rm PBH}} + 1} \cos\theta, \quad (3.3)$$

where $\theta$ is the angle between the direction of NCDM momentum and normal vector of the event horizon. Then the total energy radiated in the form of NCDM by all PBHs is given by the Stefan-Boltzmann equation multiplies with the number of PBHs,

$$\frac{d\rho_{\rm NCDM}}{dt} = L \cdot 4\pi r_S^2 \cdot n_{\rm PBH}(t). \quad (3.4)$$

By relating Eqs. (3.2) and (3.4), we get the differential PSD for the NCDM

$$\frac{d\mathcal{F}_{\rm NCDM}}{dt} = \frac{\mathcal{G}\pi r_S^2}{e^{xT/T_{\rm PBH}} + 1} n_{\rm PBH}(t). \quad (3.5)$$

For the convenience of our calculation, we change the PSD of NCDM to be the function of temperature and do the integral from $T_i$ to some later time $T$

$$\mathcal{F}_{\rm NCDM}(x,T) = \int_{T_i}^{T} dT' \frac{-1}{\mathcal{H}(T')T'} \frac{\mathcal{G}\pi r_S^2}{e^{xT'/T_{\rm PBH}} + 1} \frac{\rho_r(T_i)}{M_{\rm PBH,0}} \beta \left(\frac{T'}{T_i}\right)^3 \propto M_{\rm PBH,0}^{\frac{1}{2}}, \quad (3.6)$$

where we use Eq. (2.2) to rewrite $n_{\rm PBH}(t)$ with parameter $\beta$. In the left panel of Fig. 2, we plot the PSD of NCDM produced from PBH Hawking radiation with fixed $\beta$. As we can see, NCDM PSD is slightly proportional to PBH initial mass. As a result, those produced by a heavier PBH can populate in the higher energy region and therefore should cause greater deviation in the matter power spectrum. Note that the integration limit should start from the temperature when the PBH is formed given by Eq. (2.1) and end sufficiently late when all PBHs have been evaporated. In this way, we can properly account for all NCDMs emitted throughout the PBH's lifetime.

By definition, the number density of NCDM is

$$\begin{aligned} \frac{dn_{\rm NCDM}}{dt} &= g_{\rm NCDM} \int \frac{d^3\vec{p}}{(2\pi)^3} \frac{d\mathcal{F}_{\rm NCDM}}{dt} \\ &= \frac{g_{\rm NCDM}}{2\pi} \mathcal{G} r_S^2 n_{\rm PBH} T^3 \int_0^\infty dx \frac{x^2}{e^{xT'/T_{\rm PBH}} + 1} \\ &= \frac{3\zeta(3)}{4\pi} \frac{\mathcal{G}}{16\pi^2} g_{\rm NCDM} T_{\rm PBH} n_{\rm PBH} \left(\frac{T}{T'}\right)^3, \end{aligned} \quad (3.7)$$

where $T'$ will be integrated into the subsequent step. Also, in the final step, we perform the $x$ integral and use a convenient relation: $r_S T_{\rm PBH} = 1/4\pi$ to simplify. Again, we change to temperature integral

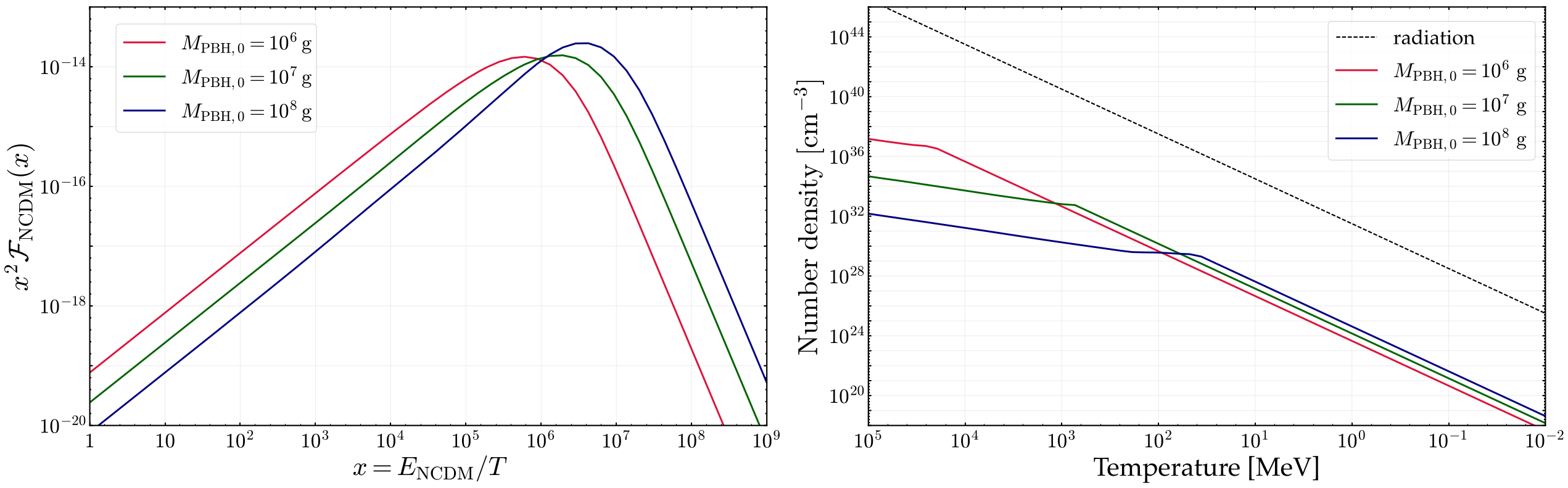

FIG. 2. Left panel: PSD of DM emitted from PBH Hawking radiation with three different choices of PBH mass. The parameter $\beta$ is fixed to $10^{-13}$. Right panel: Temperature evolution of DM number density. Again with three different PBH masses and the fixed $\beta = 10^{-13}$ in each case. The evolution experiences two stages: first when the PBH is not yet evaporated, the DM number density decreases as $T$, much slower than that of radiation. Upon PBH evaporation, there is a little boost in the number density, but after that, there is a lack of production source, and the number density dilutes as the usual $T^3$.

$$n_{\mathrm{NCDM}}(T \geq T_{\mathrm{eva}}) = \frac{3\zeta(3)}{4\pi}\frac{\mathcal{G}}{16\pi^2} g_{\mathrm{NCDM}} \int_{T_i}^{T} dT' \frac{-1}{\mathcal{H}(T')T'} T_{\mathrm{PBH}} \frac{\rho_r(T_i)}{M_{\mathrm{PBH},0}} \beta \left(\frac{T'}{T_i}\right)^3 \left(\frac{T}{T'}\right)^3$$
$$= \frac{3\zeta(3)}{4\pi}\frac{\mathcal{G}}{16\pi^2} g_{\mathrm{NCDM}} \frac{\beta\rho_r(T_i)}{M_{\mathrm{PBH},0}T_i^3} T^3 \int_{T_i}^{T} dT' \frac{-1}{\mathcal{H}(T')T'} T_{\mathrm{PBH}}. \tag{3.8}$$

Since PBH evaporation can be considered as instantaneous, before PBH evaporation, all the PBH related quantities, such as mass, temperature, and radius, can be approximated as constants. An interesting observation is that the integrand of Eq. (3.8) is proportional to $T'^{-3}$ with the above approximation. After integrating over $T'$ and canceling out with $T^3$ dependence outside the integral, we get a simple relation: $n_{\mathrm{NCDM}} \propto T$, recognizing the fact that even the population of NCDM is diluted with the expansion of the Universe, more and more are still being produced by the PBH. The right panel of Fig. 2 shows the NCDM number density evolution with three different PBH masses. Before evaporation, the number density redshifts much slower than that of radiation because of the aforementioned production feature. Then, upon evaporation, $T_{\mathrm{PBH}}$ raises substantially, causing a rapid production of NCDM at the very end of the PBH lifetime. After all PBHs have evaporated, there is no source for new NCDM to be produced, thus its number density follows the usual $T^{-3}$ due to expansion,

$$n_{\mathrm{NCDM}}(T < T_{\mathrm{eva}}) = \left(\frac{T}{T_{\mathrm{eva}}}\right)^3 n_{\mathrm{NCDM}}(T_{\mathrm{eva}}). \tag{3.9}$$

A sanity check can be done by considering the particle emitting rate for a single PBH within an energy interval [54]

$$\frac{dN_{\mathrm{NCDM}}}{dtdE} = \frac{g_{\mathrm{NCDM}}}{2\pi}\frac{\Gamma_{\mathrm{NCDM}}(E, M_{\mathrm{PBH},0})}{e^{E/T_{\mathrm{PBH}}}+1}, \tag{3.10}$$

where $\Gamma_{\mathrm{NCDM}} = 27E^2M_{\mathrm{PBH},0}^2/M_{\mathrm{pl}}^4$ is the graybody factor in the high energy geometrical-optics limit. By assuming the mass and temperature of PBH to be constants and performing the energy integral, we get the number of NCDM per unit time

$$\frac{dN_{\mathrm{NCDM}}}{dt} = \frac{3\zeta(3)}{4\pi}\frac{27}{64\pi^2} g_{\mathrm{NCDM}} T_{\mathrm{PBH}}. \tag{3.11}$$

If we consider the whole population of a PBH and focus on $T = T'$, then we reproduce the result obtained in Eq. (3.7) with a factor of 2 difference due to the fact that high energy geometrical-optics limit ignores the small spin-dependent low-energy suppression of the spectrum.

The energy density of NCDM can be derived in a similar fashion as the number density,

$$\frac{d\rho_{\mathrm{NCDM}}}{dt} = g_{\mathrm{NCDM}} \int \frac{d^3\vec{p}}{(2\pi)^3} E \frac{d\mathcal{F}_{\mathrm{NCDM}}}{dt}$$
$$= \frac{g_{\mathrm{NCDM}}}{2\pi} \mathcal{G} r_S^2 n_{\mathrm{PBH}} T^4 \int_0^\infty dx \frac{x^3}{e^{xT'/T_{\mathrm{PBH}}}+1}$$
$$= \frac{7\pi^3}{240}\frac{\mathcal{G}}{16\pi^2} g_{\mathrm{NCDM}} T_{\mathrm{PBH}}^2 n_{\mathrm{PBH}} \left(\frac{T}{T'}\right)^4, \tag{3.12}$$

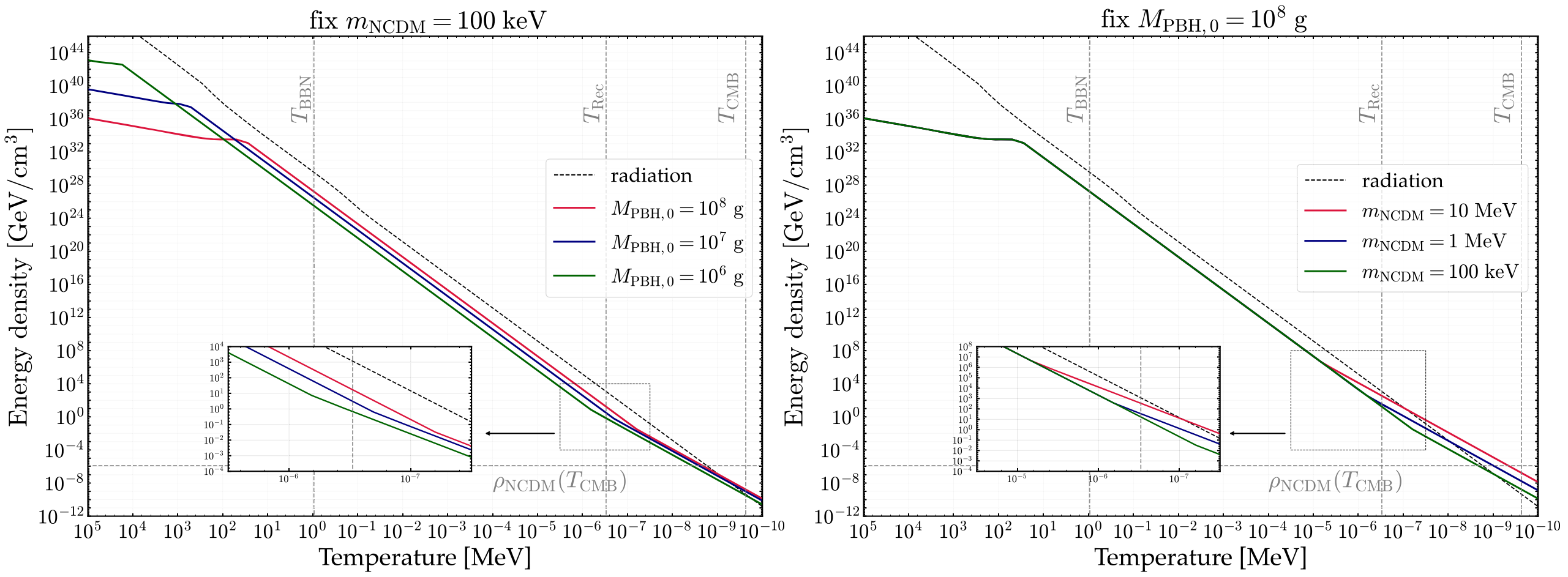

FIG. 3. Left panel: NCDM energy density evolution with fixed NCDM mass. For heavier PBH, the production phase lasts longer than that of lighter ones, which means NCDM will have less time to cool off and eventually could still be hot when structure formation takes place. Right panel: NCDM energy density evolution with fixed PBH mass. Since the production phase is independent of NCDM mass, the evolution history only deviates when NCDM becomes matterlike. Heavier NCDMs turn nonrelativistic earlier than lighter ones, and therefore becomes part of the usual CDM causing no damage on the structure formation. In both panels, we fix $\beta = 10^{-14}$.

$$\begin{aligned}\rho_{\rm NCDM}(T > T_{\rm eva}) &= \frac{7\pi^3}{240}\frac{\mathcal{G}}{16\pi^2} g_{\rm NCDM} \int_{T_i}^{T} dT' \frac{-1}{\mathcal{H}(T')T'} T_{\rm PBH}^2 \frac{\rho_r(T_i)}{M_{\rm PBH,0}} \beta \left(\frac{T'}{T_i}\right)^3 \left(\frac{T}{T'}\right)^4 \\ &= \frac{7\pi^3}{240}\frac{\mathcal{G}}{16\pi^2} g_{\rm NCDM} \frac{\beta\rho_r(T_i)}{M_{\rm PBH,0}T_i^3} T^4 \int_{T_i}^{T} dT' \frac{-1}{\mathcal{H}(T')T'^2} T_{\rm PBH}^2. \end{aligned} \tag{3.13}$$

As shown above, before PBH evaporation, the energy density of NCDM scales as $T$ until the assumption, $T_{\rm PBH} \sim$ Const., becomes invalid. In Fig. 3, we plot the evolution of NCDM energy density with photon temperature and fix $\beta = 10^{-14}$. For each set of parameters, NCDM energy density goes through three stages of evolution:

(i) Before PBHs all evaporate away: This is the production phase of NCDM. Despite being produced, the energy density still dilutes with the expansion of the Universe. However, it decreases linearly with $T$, which is much slower than the radiation ($\rho_r \propto T^4$), as described in the above paragraph and shown in Eq. (3.13). Upon PBH evaporation, there is a small upturn in the energy density corresponding to a rapid production phase when the PBH is about to evaporate entirely.

(ii) After PBH evaporation: NCDM loses its production source but since we have the condition: $E_{\rm NCDM} \sim T_{\rm PBH} \gg m_{\rm NCDM}$, NCDM is ultrarelativistic and its energy density scales as $T^4$,

$$\rho_{\rm NCDM}(T_{\rm NR} < T < T_{\rm eva}) = \left(\frac{T}{T_{\rm eva}}\right)^4 \rho_{\rm NCDM}(T_{\rm eva}). \tag{3.14}$$

(iii) Become matterlike: Finally, as the energy decreases with the expansion of Universe, NCDM becomes matterlike around $E_{\rm NCDM} \sim m_{\rm NCDM}$. We define $T_{\rm NR}$ as the temperature when NCDM is nonrelativistic, which can be estimated by the conservation of entropy in the thermal plasma,

$$\begin{aligned} &g_{*S}(T_{\rm eva})T_{\rm eva}^3 a_{\rm eva}^3 = g_{*S}(T_{\rm NR})T_{\rm NR}^3 a_{\rm NR}^3 \\ &\Rightarrow T_{\rm NR} = \left(\frac{g_{*S}(T_{\rm eva})}{g_{*S}(T_{\rm NR})}\right)^{\frac{1}{3}} T_{\rm eva}\frac{a_{\rm eva}}{a_{\rm NR}} \simeq \left(\frac{g_{*S}(T_{\rm eva})}{g_{*S}(T_{\rm NR})}\right)^{\frac{1}{3}} \frac{T_{\rm eva}}{T_{\rm PBH}} m_{\rm NCDM} \\ &\simeq 6.15\times 10^{-9}\ {\rm MeV}\left(\frac{10^6\ {\rm g}}{M_{\rm PBH,0}}\right)^{\frac{1}{2}}\left(\frac{m_{\rm NCDM}}{1\ {\rm keV}}\right), \end{aligned} \tag{3.15}$$

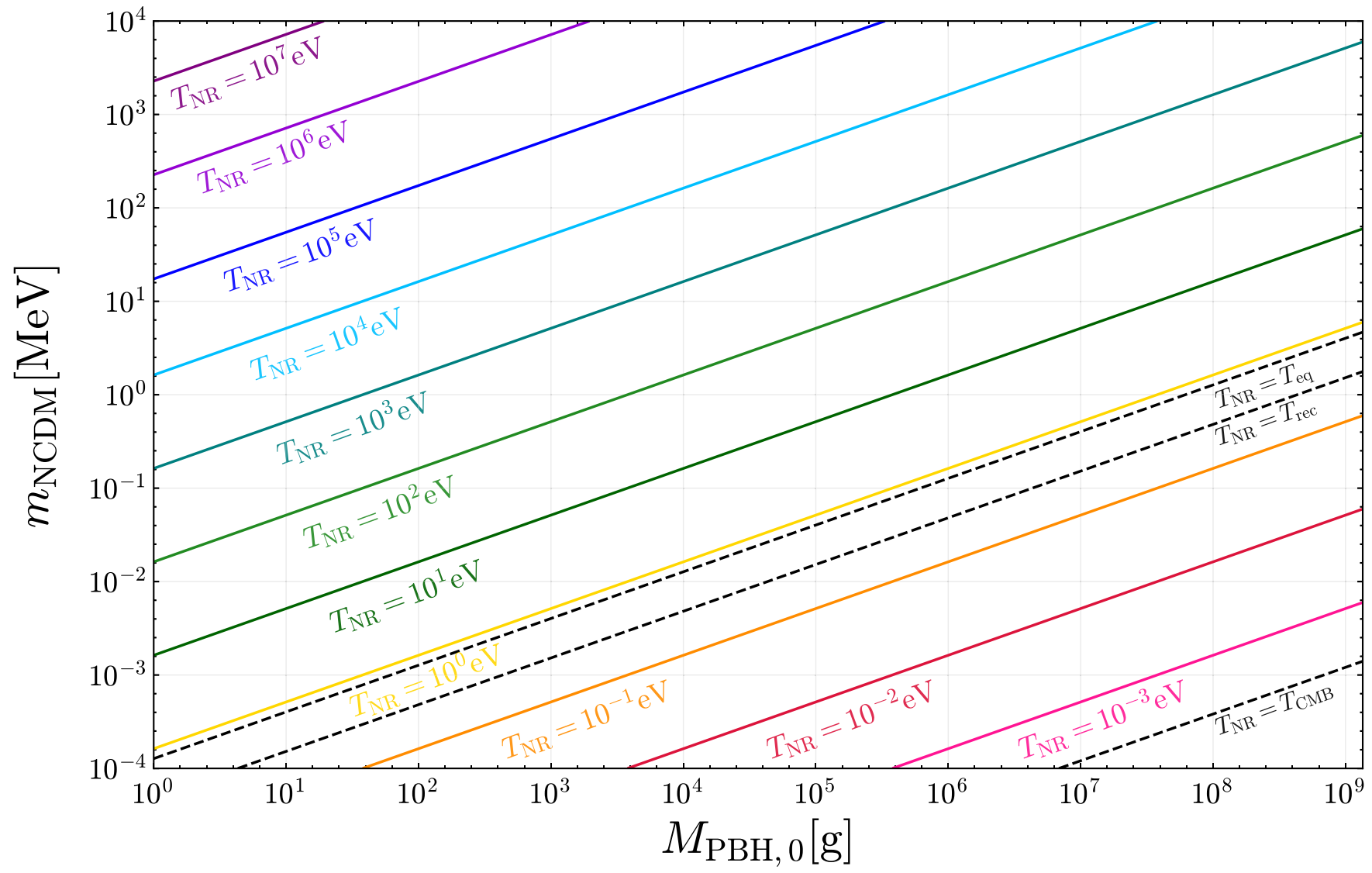

FIG. 4. Contours of temperature when NCDM becomes nonrelativistic shown in the $M_{PBH,0} - m_{NCDM}$ parameter space. According to Eq. (3.15), the slope of contours is $\frac{1}{2}$.

where in the second to last expression we use the assumption: $E_{NCDM}(T_{NR}) = a_{eva}/a_{NR} * T_{PBH} \simeq m_{NCDM}$. As we can see, the keV scale NCDMs produced by a $10^6$ g PBH with lifetime around $4 \times 10^{-10}$ sec needs an incredible amount of time to turn nonrelativistic. Before that, NCDMs are likely to smooth out small scale structure as we expected. $T_{NR}$ will plays a crucial role in analyzing NCDM effect on structure formation. In Fig. 4, we shows the $T_{NR}$ contours in the mass of NCDM and PBH parameter space. Different $T_{NR}$ results in distinct phenomenology, we will scrutinize the effects in the following section. After becoming matterlike, the NCDM energy density scales as

$$\rho_{NCDM}(T < T_{NR}) = \left(\frac{T}{T_{NR}}\right)^3 \left(\frac{T_{NR}}{T_{eva}}\right)^4 \rho_{NCDM}(T_{eva}). \tag{3.16}$$

In the left panel of Fig. 3, we fix the NCDM mass to be 100 keV and vary the PBH mass. A crucial point that will be emphasized multiple times throughout the paper is the following: a heavier PBH evaporates slower, thus elongates the production phase of NCDM; on the other hand, even though NCDMs from lighter PBH are more energetic in the beginning, they have much more time to cool off at the same time. At the time of structure formation, a NCDM from a heavier PBH is more likely to be energetic and cause the suppression of small scale structure. In the right panel of Fig. 3, we fix the PBH mass to be $10^8$ g and vary the NCDM mass. Now as PBH mass and $\beta$ are fixed, the production phase is consequently fixed and the only question is when NCDM becomes nonrelativistic. Heavier NCDM turns nonrelativistic earlier than the lighter one, therefore constituting a larger portion in the DM relic abundance today.

The relic abundance of the NCDM from PBH can be computed as

$$\begin{aligned}\Omega_{NCDM}h^2 &= \frac{\rho_{NCDM}(T_{CMB})}{\rho_{crit}} h^2 \\ &\simeq 0.12 \left(\frac{g_*(T_{eva})}{88.92}\right)^{\frac{1}{3}} \left(\frac{M_{PBH,0}}{10^6\ \mathrm{g}}\right)^{\frac{1}{2}} \left(\frac{m_{DM}}{1\ \mathrm{keV}}\right) \\ &\quad \times \left(\frac{\beta}{4.05 \times 10^{-9}}\right).\end{aligned} \tag{3.17}$$

From Eq. (3.17), we can derive another constraint on $\beta$ by the overproduction of NCDM, which has a different parametric dependence than the early PBH dominant constraint derived in Eq. (2.9). Depending on the chosen parameters, two constrains could be complimentary to each other. In the next section, however, we will show that these two bounds are much weaker for sub-GeV scale NCDM when confronted with the precise matter spectrum measurement.

## IV. CONSTRAINT FROM STRUCTURE FORMATION

It is well known that the NCDM is constrained by the ΛCDM model due to its free-streaming effect that smooths out the density fluctuation and suppress the matter power spectrum in the nonlinear regime. In this work, we take the PBH's mass and the abundance $\beta$ to be the free parameters and use Eq. (3.6) to generate DM PSD, which is then fed into the linear Boltzmann solver code CLASS [55] to obtain the matter power spectrum as well as the CMB spectrum.

As we discuss in the previous section, a heavier PBH evaporates slower and is able to constantly produce NCDM until a much later time. As a consequence, in Fig. 3, NCDMs from heavier PBH retain larger energy at structure formation compared to those from lighter ones and lead to larger suppression in the matter power spectrum. We can see this effect clearly in Fig. 1, in which the total DM relic abundance is fixed to $\Omega_{\rm DM}^{\rm tot}h^2 = 0.12$ [56] and we choose $m_{\rm NCDM} = 1$ MeV and $\beta = 10^{-13}$. Also, we show $\Omega_{\rm NCDM}h^2$, the relic abundance of NCDM. Compared to $\Omega_{\rm DM}^{\rm tot}h^2$, only about 1 percent of NCDM is enough to smooth out the structure and cause significant deviation from $\Lambda$CDM prediction. We follow [57] for data processing and do the $\Delta\chi^2$ test

$$\Delta\chi^2 \equiv \chi^2_{\rm NCDM} - \chi^2_{\rm Std} = \sum_i \left[ \frac{(P_k^{\rm NCDM}(k_i) - P_k^{\rm Obs}(k_i))^2}{\left(\Delta P_k^{\rm Obs}(k_i)\right)^2 + \left(\frac{d}{dk} P_k^{\rm Obs}(k_i) \times \Delta k_i\right)^2} - \frac{(P_k^{\rm Std}(k_i) - P_k^{\rm Obs}(k_i))^2}{(\Delta P_k^{\rm Obs}(k_i))^2 + \left(\frac{d}{dk} P_k^{\rm Obs}(k_i) \times \Delta k_i\right)^2} \right], \quad (4.1)$$

where $i$ sums over all available data points, and NCDM, Std, and Obs are the matter power spectrum simulated in the presence of NCDM from a PBH, the standard $\Lambda$CDM from CLASS, and observations respectively. Thanks to the precise measurements (especially SDSS) on small scale, we are able to put strict constraint on the PBH by the NCDM it produces.

In the upper row of Fig. 5, we show the constraint on PBH mass and $\beta$ parameter space with NCDM mass ranges between 0.1 keV and 10 GeV. Above the black dashed line, in the gray shaded region, PBH production is too efficient and its energy density becomes larger than that of radiation before evaporation causing a PBH domination that violates our assumption in Eq. (2.9). The colored dashed lines in the upper right panel are the upper bounds from NCDM overproduction, $\Omega_{\rm NCDM}h^2 > 0.12$, which we consider as a hard cutoff for the abundance of PBHs. For a light NCDM with mass around the keV range, this bound is much weaker than the PBH domination. However, as Eq. (3.17) shows, the relic density depends on NCDM mass, and the overproduction constraint is complementary to PBH domination when NCDM is heavier than $\mathcal{O}(10\ \text{MeV})$. The colored solid lines are the constraint from the matter power spectrum. Every solid colored line, joining with the black dashed line, rules out the region above. For the lightest two cases of NCDM, we use dotted lines to show the matter power spectrum constraint, which extends to the PBH dominated region. We see that every curve has a small upturn at $T_{\rm eva} \sim T_{\rm QCD}$ corresponding to the discussion below Eq. (2.4) where the QCD phase transition suddenly reduces the relativistic degree of freedom, increases the evaporating temperature, and in the end slightly relaxes the constraint. While in the lower row, we translate the matter power spectrum constraint into the constraint on the fraction of NCDM from PBH in the population of total DM, $f_{\rm NCDM} = \Omega_{\rm NCDM}/\Omega_{\rm DM}^{\rm tot}$. For light NCDMs, even a tiny fraction of NCDM can be constraint. However, for heavier NCDM, the constraint becomes weaker as it turns nonrelativistic long before structure formation and eventually is replaced by the NCDM overproduction bound. In the following subsections, we categorize our discussion based on when NCDM becomes nonrelativistic, which results in distinctive physics.

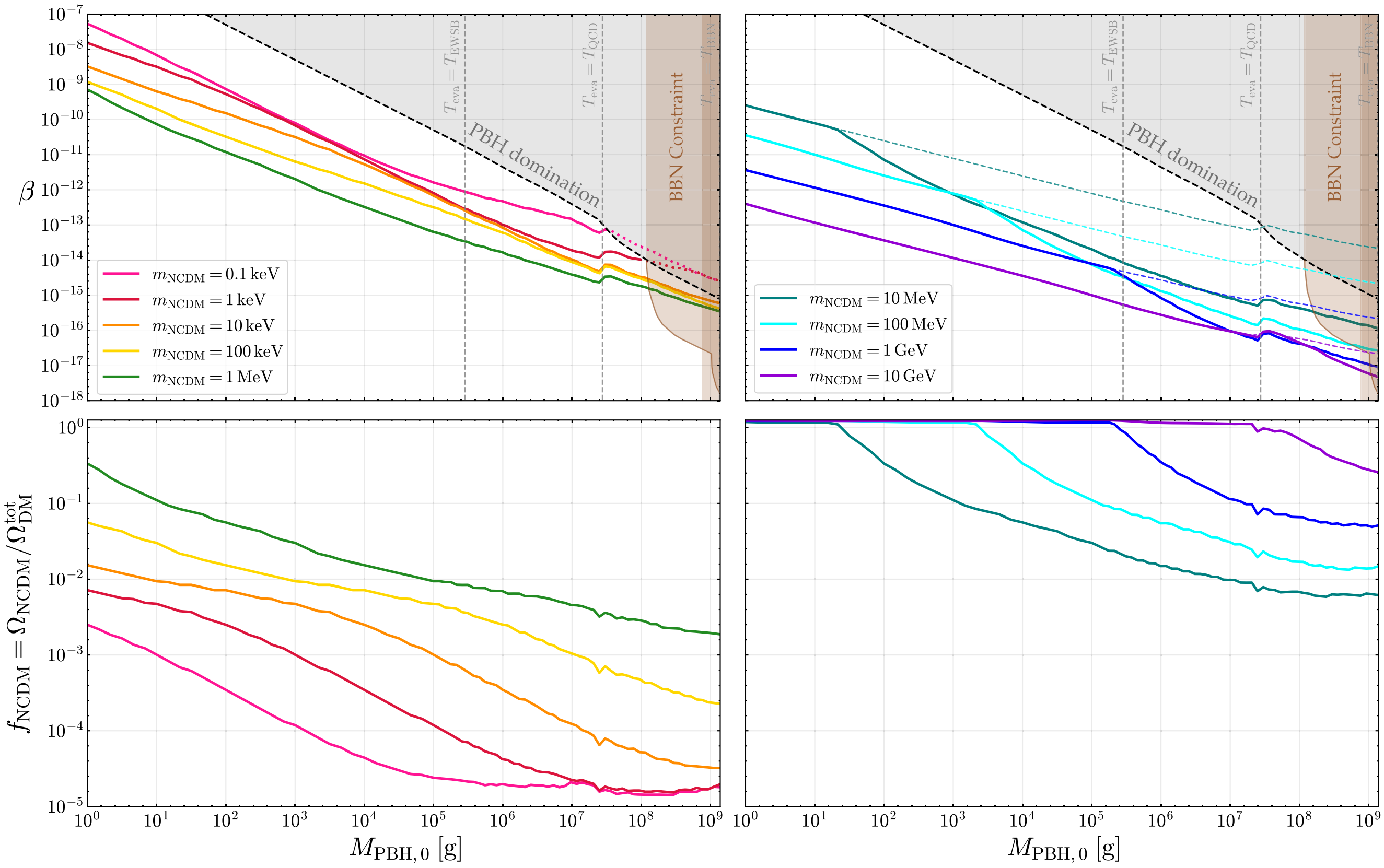

FIG. 5. Upper row: matter power spectrum constraint on the PBH masses and $\beta$ parameter space. Colored lines are constraints derived from different masses of NCDM. Black dashed line indicates the boundary between radiation domination (below) and PBH domination (above). We also show the BBN constraint [28,29] in brown. Lower row: constraint on fraction of NCDM produced by PBH, translated from upper row.

### A. NCDMs remain relativistic at recombination: $T_{\rm NR} < T_{\rm Rec}$

We start with the scenario where NCDMs are still relativistic at recombination. In this region, NCDMs behave as dark radiation and the constraint is mainly from $\Delta N_{\rm eff}$. We will further divide it into two subsets based on whether NCDMs remain relativistic today.

(i) NCDMs remain relativistic today: This condition can only be achieved in the combination of light NCDM and heavy PBH as shown in the bottom-right corner of Fig. 4 by the black-dashed line: $T_{\rm NR} = T_{\rm CMB}$. In Fig. 5, only the lightest two cases of NCDM mass can fulfill the requirement. In the upper left panel, at the far right of pink ($m_{\rm NCDM} = 0.1$ keV) and red ($m_{\rm NCDM} = 1$ keV), they join together when $M_{\rm PBH,0} \gtrsim 5 \times 10^8$ g; while in the lower panel, they saturate into a constant value, independent of $m_{\rm NCDM}$ and $M_{\rm PBH,0}$. Being dark radiation, NCDMs are contributed to the extra relativistic degrees of freedom, which can be estimated by adopting $\Delta N_{\rm eff} < 0.3$ at 95% in [45]

$$
\begin{aligned}
\Delta N_{\rm eff} &= \frac{\rho_{\rm NCDM}(T_{\rm rec})}{\rho_\nu(T_{\rm rec})} = \frac{\rho_{\rm NCDM}(T_{\rm CMB})\left(\frac{T_{\rm rec}}{T_{\rm CMB}}\right)^4}{\rho_\nu(T_{\rm CMB})\left(\frac{T_{\rm rec}}{T_{\rm CMB}}\right)^4} \\
&= \frac{\Omega_{\rm NCDM}}{\Omega_\nu} \\
\Rightarrow f_{\rm NCDM} &= \frac{\Omega_\nu \times \Delta N_{\rm eff}}{\Omega_{\rm DM}^{\rm tot}} \sim \frac{\Omega_\gamma \times \Delta N_{\rm eff}}{\Omega_{\rm DM}^{\rm tot}} \\
&\sim \mathcal{O}(10^{-5}),
\end{aligned}
\tag{4.2}
$$

where in the second to last approximation, we approximate the neutrino relic density to that of photon based on the argument that neutrinos are treated as massless in ΛCDM and they contribute similarly to photons in entropy density. This fraction could be thought of as the minimum required fraction of NCDM to cause visible impact on the matter power spectrum. We could also expect this outcome from Fig. 3, as it indicates that the mass of NCDM is only relevant at determining the time NCDM becomes nonrelativistic. Therefore, if NCDMs are radiationlike until today, with the same $M_{\rm PBH,0}$ and $\beta$, they share the same evolution history and result in the same constraint.

(ii) NCDMs are nonrelativistic today: We can reach this region in parameter space by either decreasing the PBH mass or increasing the NCDM mass. Similar to the above argument, NCDMs behave as dark radiation at recombination and increase the $N_{\rm eff}$. However, the evolution path afterward is different than the previous one as now NCDMs turn nonrelativistic midway between recombination and today. We need to take into account the conversion at $T_{\rm NR}$, thus the contribution to $\Delta N_{\rm eff}$ is

$$
\begin{aligned}
\Delta N_{\rm eff} &= \frac{\rho_{\rm NCDM}(T_{\rm rec})}{\rho_\nu(T_{\rm rec})} \\
&= \frac{\rho_{\rm DM}(T_{\rm CMB})\left(\frac{T_{\rm NR}}{T_{\rm CMB}}\right)^3\left(\frac{T_{\rm rec}}{T_{\rm NR}}\right)^4}{\rho_\nu(T_{\rm CMB})\left(\frac{T_{\rm rec}}{T_{\rm CMB}}\right)^4} \\
&= \frac{\Omega_{\rm NCDM}}{\Omega_\nu}\frac{T_{\rm CMB}}{T_{\rm NR}} \\
\Rightarrow f_{\rm NCDM} &= \frac{\Omega_\nu \times \Delta N_{\rm eff}}{\Omega_{\rm DM}^{\rm tot}}\frac{T_{\rm NR}}{T_{\rm CMB}} \\
&\propto m_{\rm NCDM} M_{\rm PBH,0}^{-\frac{1}{2}}.
\end{aligned}
\tag{4.3}
$$

As shown in the lower-left panel of Fig. 5, each solid line exhibits a region where the fraction is inversely proportional to the square root of PBH masses, and the separations between lines are proportional to the NCDM mass. At the exact same range of $M_{\rm PBH,0}$, we see three overlaps in the upper left panel, they are *(i)* $3 \times 10^2$ g $\lesssim M_{\rm PBH,0} \lesssim 10^4$ for pink and red; *(ii)* $2 \times 10^4$ g $\lesssim M_{\rm PBH,0} \lesssim 10^6$ for red and orange and *(iii)* $4 \times 10^6$ g $\lesssim M_{\rm PBH,0} \lesssim 10^9$ for orange and yellow. This can be understood from Eq. (3.17) that if PBH mass is fixed, we have $\beta \propto \Omega_{\rm NCDM}/m_{\rm NCDM}$. As a result, because of the same $\Omega_{\rm NCDM}/m_{\rm NCDM}$ ratio across different lines in their respective overlap region, it gives rise to the same constraint in $\beta$.

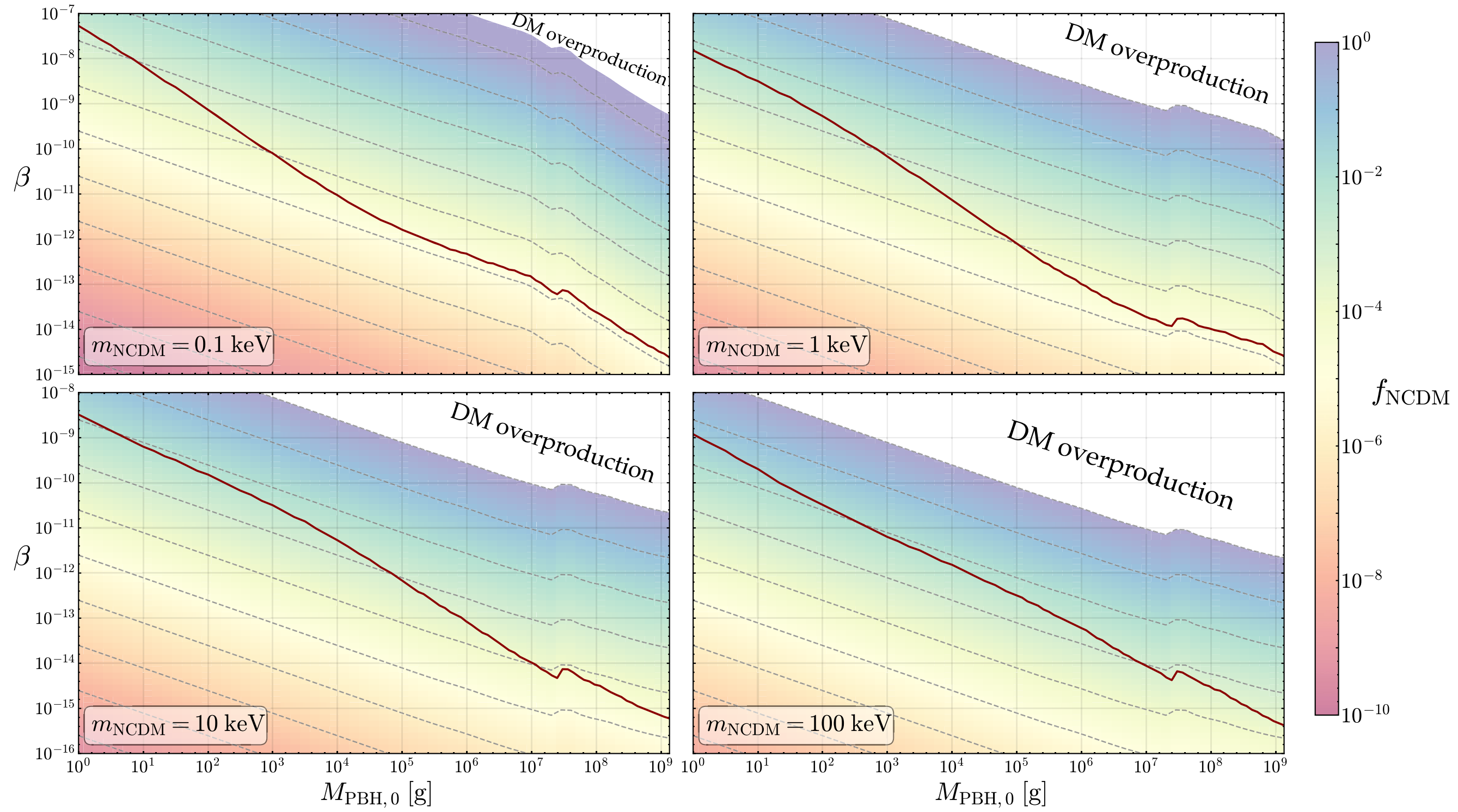

FIG. 6. Matter power spectrum on PBH abundance with NCDM fraction contours for the lightest four NCDM masses. Dark red lines are constraints from upper left panel of Fig. 5, which exclude the region above. Across the whole parameter space, the LSS constraint is able to limit the fraction of NCDM from PBH to the whole NCDM population to percent level.

For the lightest four cases, we show in Fig. 6 the $f_{\rm NCDM}$ contour on the parameter space of $M_{\rm PBH,0} - \beta$. Dark red lines, taken from the upper left panel of Fig. 5, rule out the parameter space above it. As we can see, across the parameter space, only a small fraction of light NCDMs, produced via PBH evaporation, which possess large free-streaming lengths, are required to effectively suppress the formation of small-scale structures.

### B. Nonrelativistic before recombination, $T_{\rm NR} > T_{\rm Rec}$

For NCDMs to turn nonrelativistic slightly earlier than recombination happens, their energy density redshifts slower than that of radiation and increases the energy density ratio between these two fluids. We show in Fig. 7, the marginal case of the allowed energy density ratio to photon at $T = 0.3$ eV translated from the upper left panel of Fig. 5. When they move from a heavy PBH to a lighter one, every curve has a "$V$" shape, where the energy density ratio decreases to a universal minimum then increases monotonically.

We first address the right part of the "$V$." In this region, NCDMs still act as radiation when recombination happens as outlined in the previous subsection and are subjected to the $\Delta N_{\rm eff}$ constraint. Their energy density at recombination can be estimated as

$$\rho_{\rm NCDM}(T_{\rm rec}) = \rho_{\rm NCDM}(T_{\rm CMB})\left(\frac{T_{\rm NR}}{T_{\rm CMB}}\right)^3\left(\frac{T_{\rm rec}}{T_{\rm NR}}\right)^4 \propto \frac{1}{T_{\rm NR}} \propto m_{\rm NCDM}^{-1} M_{\rm PBH,0}^{\frac{1}{2}}, \quad (4.4)$$

while the upturns of the curve match the PBH mass when the fraction constraints in the lower panel of Fig. 5 stop following the $M_{\rm PBH,0}^{-\frac{1}{2}}$ dependence in Eq. (4.3). Then in the left part of the "$V$," NCDMs become nonrelativistic before recombination and their energy density scales as

$$\rho_{\rm NCDM}(T_{\rm rec}) = \rho_{\rm NCDM}(T_{\rm CMB})\left(\frac{T_{\rm NR}}{T_{\rm CMB}}\right)^3 \propto T_{\rm NR}^3 \propto M_{\rm PBH,0}^{-\frac{3}{2}}. \quad (4.5)$$

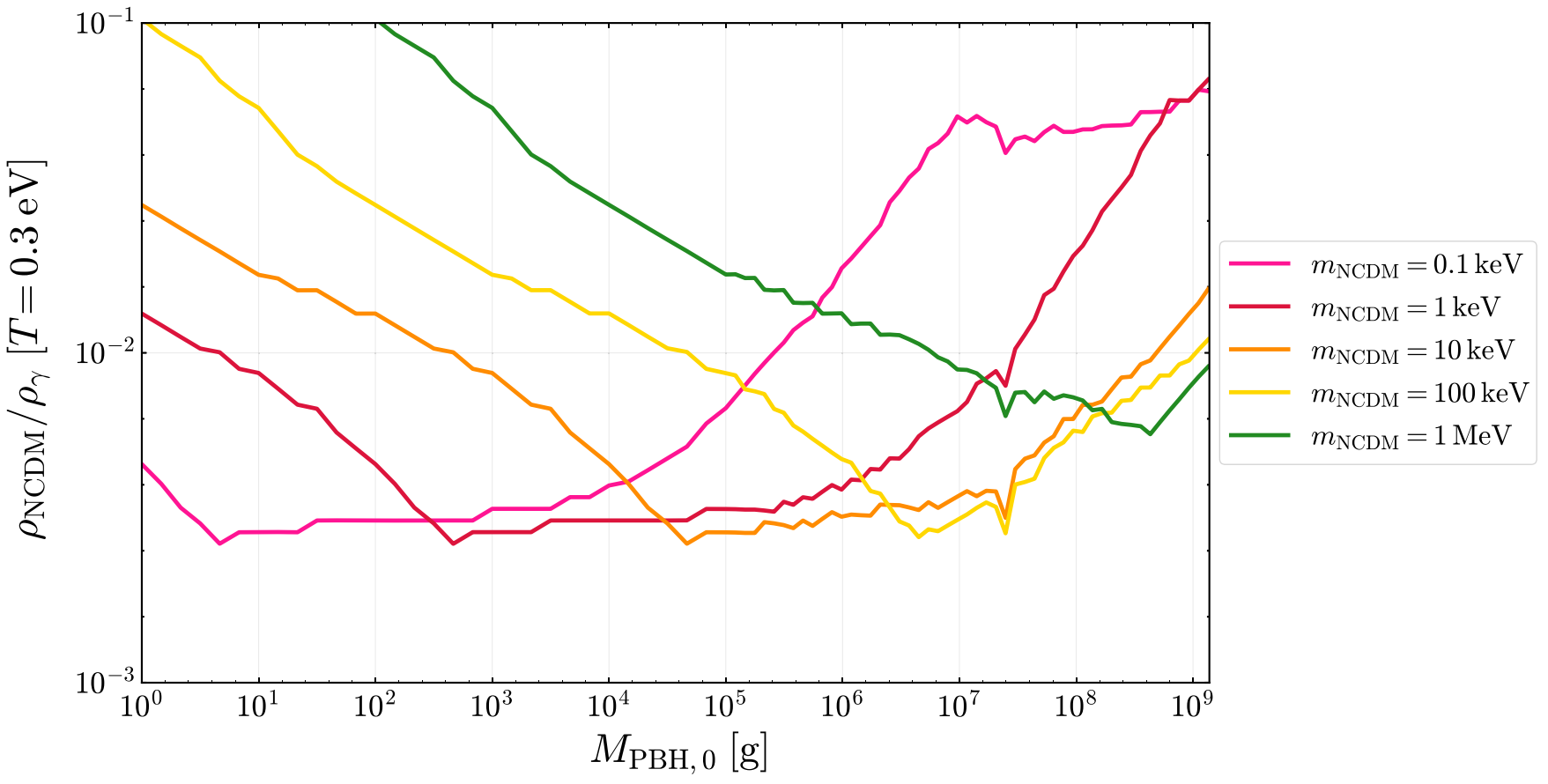

FIG. 7. Energy density ratio between NCDM and the photon at $T = 0.3$ eV translated from the upper row of Fig. 5. The allowed ratio first decreases then increases as the PBH becomes heavier. The declined part is because NCDM turns nonrelativistic before recombination, making its energy density redshifts slower than that of radiation and brings the ratio up. In the inclined part, NCDM is still relativistic at recombination, the heavier PBH results in larger NCDM energy density, larger $\Delta N_{\rm eff}$, and thus stronger constraint.

The physical interpretation arises from the fact that NCDMs originally possess a much greater energy compared to ordinary CDM. Although their energy decreases with redshift and NCDMs slowly become matterlike, their free-stream lengths remain larger than that of CDM. As a consequence, a larger ratio to photon energy density means these particles occupy a larger portion in the total energy budget of the Universe and result in a flatter dependence on $M_{\rm PBH,0}$ in Fig. 5.

When the mass of NCDM is around the MeV scale, we see a rapid decrease of the constraint, particularly in the light PBH region. The reason is due to the fact that the light PBH evaporates much earlier than the heavier one and heavier NCDM turns nonrelativistic earlier. As such, PBH-produced NCDM becomes regular CDM and constraint fades away as expected. Finally, the DM overproduction constraint sets in, effectively replacing the matter power spectrum constraint for lighter PBH and heavier NCDM.

### C. PBH domination

In this subsection, we comment on the scenario when the PBH came to dominate the Universe before the BBN. As we stated, Eq. (2.9) is just a soft bound since we have no observation on the pre-BBN Universe. To estimate the effect, we can use the sudden decay approximation,

$$\tau_{\rm PBH}^{-1} = \mathcal{H}_{\rm before} = \mathcal{H}_{\rm after}. \tag{4.6}$$

Since NCDM only constitutes a small amount in the total energy injection from PBH, we will ignore the difference here and assume $\rho_{\rm PBH} \sim \rho_{\rm inj}$. After PBH evaporation, the Universe is back to radiation domination with a reheating temperature $T_{\rm RH}$, we can relate the Hubble parameter right before and right after the evaporation and get

$$\rho_{\rm PBH}(T_{\rm eva}) = M_{\rm PBH} n_{\rm PBH}(T_{\rm eva}) = M_{\rm PBH} n_{\rm PBH}(T_i)\left(\frac{T_{\rm eva}}{T_i}\right)^3$$
$$= \beta\rho_\gamma(T_i)\left(\frac{T_{\rm eva}}{T_i}\right)^3 = \rho_\gamma(T_{\rm RH})$$
$$\Rightarrow T_{\rm RH} = \left(\frac{g_*(T_{\rm eva})}{g_*(T_{\rm RH})}\frac{\beta}{\beta_c}\right)^{\frac{1}{4}} T_{\rm eva}, \tag{4.7}$$

where $\beta_c$ is given by Eq. (2.9). Again, as the Universe expands, we assume NCDM turns nonrelativistic when $E_{\rm NCDM} \sim m_{\rm NCDM}$, from Eq. (3.15) we see that $T_{\rm NR}$ is increased by the same factor. A higher $T_{\rm NR}$ means NCDM becomes matterlike earlier and potentially weakens the constraint. However, from Eq. (3.6), we see that NCDM PSD is linearly proportional to $\beta$, which is a stronger dependency than Eq. (4.7). Even though increasing $\beta$ makes NCDM nonrelativistic earlier, it also increase the amount of NCDM more significantly.

Finally, we consider the effect on the free-streaming length of NCDM when PBH domination happens. We start with the definition of free-streaming length

$$\lambda = \int_{a_0}^{a_{\rm rec}} da \frac{1}{a^2 H}, \tag{4.8}$$

where $a_0$ is the scale factor when a particular NCDM is produced. After all PBHs evaporate, the Universe goes back to radiation domination, thus we can separate the above integration into two parts

$$\lambda = \int_{a_0}^{a_{\rm eva}} da \frac{1}{a^2 H} + \int_{a_{\rm eva}}^{a_{\rm rec}} da \frac{1}{a^2 H}, \tag{4.9}$$

the energy density of PBH dominates the Hubble parameter in the first term, while radiation dominates the second. We are going to argue that the first term is actually negligible compared to the second term. Therefore, even with PBH domination, the change in the free-streaming length is tiny. The first term can be calculated as

$$\lambda_{\rm PBH} = \int_{a_0}^{a_{\rm eva}} da \frac{1}{a^2 H} = \int_{a_0}^{a_{\rm eva}} da \frac{1}{a^2 \sqrt{\frac{8\pi G}{3}\beta\rho_r(T_i)\left(\frac{a_i}{a}\right)^3}}$$

$$= \sqrt{\frac{3}{8\pi G\beta\rho_r(T_i)a_i^3}} \int_{a_0}^{a_{\rm eva}} da a^{-\frac{1}{2}}$$

$$= \sqrt{\frac{3a_{\rm eva}}{2\pi G\beta\rho_r(T_i)a_i^3}}, \tag{4.10}$$

where we assume $a_{\rm eva} \gg a_0$, and the second term is

$$\lambda_r = \int_{a_{\rm eva}}^{a_{\rm rec}} da \frac{1}{a^2 H} = \int_{a_{\rm eva}}^{a_{\rm rec}} da \frac{1}{a^2 \sqrt{\frac{8\pi G}{3}\rho_r(T_i)\left(\frac{a_i}{a}\right)^4}}$$

$$= \sqrt{\frac{3}{8\pi G\rho_r(T_i)a_i^4}} a_{\rm rec}, \tag{4.11}$$

here we also assume $a_{\rm rec} \gg a_{\rm eva}$. The ratio between these two terms is

$$\frac{\lambda_{\rm PBH}}{\lambda_r} = \frac{2\sqrt{\frac{a_{\rm eva}}{\beta a_i^3}}}{\frac{a_{\rm rec}}{a_i^2}} = 2\sqrt{\frac{a_{\rm eva}a_i}{\beta a_{\rm rec}^2}}. \tag{4.12}$$

Since we have $a_{\rm rec} \gg a_{\rm eva} \gg a_i$, we expect that our constraint in Fig. 5 is likely to apply in the case of PBH domination.

## V. CONCLUSION

In this work, we explore the scenario where part of the DM is produced by light PBH Hawking radiation while the remaining part is produced by some nonthermal production mechanism. We particularly focus on PBH mass range from 1 g to $10^9$ g, the parameter space people usually overlook because they evaporate even before BBN and leave no visible effect if the entropy injection upon evaporation is negligible. However, from the estimation below Eq. (1.1) as well as Eq. (3.1), the energy of DM is much greater than both its mass and the temperature of surrounding photons, thus the DM produced in such mechanism is ultrarelativistic and constitutes as NCDM. Since the interaction between NCDM and the SM sector is solely through gravity, two fluids have never established a substantial energy exchange whatsoever. In this setup, NCDMs from PBH evaporation require a considerable time to cool off before becoming the regular CDM we observe today as indicated in Eq. (3.15).

The specific period we are focusing on is the structure formation, where the effect of this gravitationally coupled NCDM is the most pronounced, as the free stream effect smooths out small structures and causes the suppression of matter power spectrum at large $k$ shown in Fig. 1. Even a percent level of NCDM is enough to create severe deviation from observations. By various measurements on the matter power spectrum, we are able to set the most stringent constraint on the abundance of light PBH.

Our main result is in Fig. 5, where we put a constraint on PBH mass and $\beta$ parameter space with different choices of NCDM masses. The constraint is particularly sensitive to the time when NCDM becomes nonrelativistic. In general, heavier PBH evaporates later leaves NCDM no time to cool off and thus results in stronger constraint. If NCDM is light, it could still be relativistic today, and the constrain would be independent of both PBH and NCDM masses due to the identical evolution before becoming nonrelativistic. For slightly heavier NCDM, it is cold today but energetic at recombination, the constraint on $\Delta N_{\rm eff}$ shows that $f_{\rm NCDM}$ would inherit a $M_{\rm PBH,0}^{-\frac{1}{2}}$ dependence. The constant ratio between $\Omega_{\rm NCDM}$ and $m_{\rm NCDM}$ manifests itself as three overlapped regions in $M_{\rm PBH,0} - \beta$ parameter space between the four lightest NCDM cases. On the other hand, for lighter PBH, NCDM has more time to lose energy. The constraint saturates from $M_{\rm PBH,0}^{-\frac{1}{2}}$ dependence due to a larger ratio between the energy density of NCDM and radiation, but later fades away since NCDM is gradually becoming regular CDM. Eventually the LSS constraint is overtaken by a DM overproduction constraint for $m_{\rm NCDM} \gtrsim 10$ GeV. In the PBH dominant case, we argue that our constraint could still be applicable since by increasing $\beta$ the enhancement on NCDM PSD is more obvious than the rise on $T_{\rm NR}$.

Our work offers a new perspective on the previously unexplored parameter space of PBHs and allows us to probe regions that would otherwise remain inaccessible through direct detection or astrophysical searches. By studying the interplay between PBH-produced NCDM and structure formation, we provide robust and complementary constraints on light PBH. The result shows the importance of cosmological surveys in probing nonthermal dark matter scenarios and such observations can place unique and stringent constraints on the properties and abundance of light PBHs.

## ACKNOWLEDGMENTS

The author thanks Yue Zhang, Douglas Tuckler, and Bryce Cyr for helpful discussions. This work is supported by a Subatomic Physics Discovery Grant (individual) from the Natural Sciences and Engineering Research Council of Canada.

## DATA AVAILABILITY

The data are not publicly available. The data are available from the authors upon reasonable request.